\documentclass[apjl]{aastex}
\shorttitle{Effects of dust feedback on vortices}
\shortauthors{Fu et al.}

\begin{document}

\title{Effects of dust feedback on vortices in protoplanetary disks}
\author{Wen Fu{$^{1,2}$}, Hui Li{$^2$}, Stephen Lubow{$^3$}, Shengtai Li{$^2$}, Edison Liang{$^1$}}
\affil{{$^1$}Department of Physics and Astronomy, Rice University, Houston, TX 77005, USA; wf5@rice.edu\\
{$^2$}Los Alamos National Laboratory, Los Alamos, NM 87545, USA\\
{$^3$}Space Telescope Science Institute, Baltimore, MD 21218, USA}

\begin{abstract}
We carried out two-dimensional high-resolution simulations to study 
the effect of dust feedback on the evolution of vortices induced by
massive planets in protoplanetary disks. Various initial dust to gas disk
surface density ratios ($0.001$ -- $0.01$) and dust particle sizes 
(Stokes number $4\times10^{-4}$ -- $0.16$) are considered. 
We found that while dust particles migrate inwards, vortices are 
very effective in collecting them. When dust density becomes 
comparable to gas density within the vortex, a dynamical 
instability is excited and it alters the coherent vorticity pattern 
and destroys the vortex. This dust feedback effect is stronger with 
higher initial dust/gas density ratio and larger dust grain. Consequently,
we found that the disk vortex lifetime can be reduced up to a factor of 10.
We discuss the implications of our findings on the survivability of 
vortices in protoplanetary disks and planet formation. 
\end{abstract}



\section{Introduction}

Vortices induced by massive planets in protoplanetary disks have 
gained interest as a possible explanation for recent submillimeter 
observations of large-scale asymmetric features in several 
transitional disk systems \citep{isella13, marel13, casassus13, 
fukagawa13, perez14}. These vortices could result from the Rossy-Wave 
instability \citep[RWI,][]{lovelace99, li00, li01} at the edge of a 
gap opened by a massive planet. They are coherent, large-scale 
density enhancement structures adjacent to the edges of a gap/hole 
made by planets, which can potentially explain observational features 
of many transitional disks \citep{andrews11, espaillat14}. 

Disk vortices are also interesting because they can efficiently 
trap dust particles which in turn can help promote planetesimal 
formation \citep{barge95, tanga96, chavanis00, johansen04, inaba06, 
rice06, lyra09a, lyra09b, meheut12, pinilla12, zhu12, birnstiel13, 
lyra13, zhu14}. However, most of the previous studies of dusty 
vortices have not included the dust feedback, i.e. back-reaction 
from dust onto the gas. Some studies that did include this 
effect in either the shearing-sheet approximation or global
disk simulations \citep{johansen04, lyra09b, meheut12} are all 
for the very early stage of vortex formation, up to only tens 
of orbits whereas the vortex could live for much longer \citep{fu14}. 
The long-term feedback by dust on vortices is still an open question.
We address this issue in this \textit{Letter} by carrying out 
high-resolution two-dimensional numerical simulations. 
In Section 2, we present the detailed set-up of our numerical simulations. 
We summarize our main results in Section 3, and discuss their 
implications in Section 4. 


\section{Numerical Setup}
In our study, the protoplanetary disks are assumed to be geometrically thin so that the hydrodynamical equations can be reduced to two-dimensional Navier-Stokes equations by considering vertically integrated quantities. We adopt an isothermal equation of state $P=c_s^2 \Sigma_g$ for the gas component in the disk with $P$, $c_s$, $\Sigma_g$ being gas pressure, sound speed, gas surface density, respectively. We treat the dust component as a pressureless fluid whose evolution is governed by conservation of mass, radial and angular momentum. The drag forces are incorporated into the momentum equations as external force terms, i.e., 

\begin{equation}
\frac{\partial \Sigma_d\mathbf{v}_d}{\partial t} +\nabla(\mathbf{v}_d\cdot\Sigma_d \mathbf{v}_d)=-\Sigma_d\nabla\Phi_{G}+\Sigma_d \mathbf{f}_d,
\label{eq:eq1}
\end{equation}
where $\Phi_{G}$ is the gravitational potential of the central star and the planet, $\Sigma_d$, $\mathbf{v}_d$ are dust fluid surface density and velocity, respectively. The drag force $\mathbf{f}_d$ between the gas and the dust is 
\begin{equation}
\mathbf{f}_d=\frac{\Omega_k}{\mathrm{St}}(\mathbf{v}_g-\mathbf{v}_d),
\end{equation}
where $\Omega_k$ is the Keplerian angular velocity, $\mathbf{v}_g$ is the gas fluid velocity, and $\rm{St}$ is the Stokes number of dust particles \citep[e.g.,][]{takeuchi02},
\begin{equation}
\rm{St}=\sqrt{\frac{\pi}{8}}\frac{\rho_p s_p \Omega_k}{\rho_g c_s}=\frac{\pi s_p \rho_p}{2\Sigma_g},
\end{equation}
where $s_p$ is the diameter of dust particle and $\rho_p$ is the internal density of dust particle, and for the second equal sign we have used the fact that $c_s=H\Omega_k$, $\Sigma_g=\sqrt{2\pi}H\rho_g$ with $H$ being disk scale height. Here we have assumed that the particle size is smaller than the mean free path of gas molecules so that dust-gas interaction is well inside the Epstein regime. To study the dust-gas joint evolution, we include the drag force also in the momentum equation for gas

\begin{equation}
\frac{\partial_t \Sigma_g\mathbf{v}_g}{\partial t} +\nabla(\mathbf{v}_g\cdot\Sigma_g \mathbf{v}_g)+\nabla P=-\Sigma_g\nabla\Phi_{G}+\Sigma_g \mathbf{f}_{\nu}-\Sigma_g \mathbf{f}_d,
\label{eq:eq2}
\end{equation}
where $\mathbf{f}_{\nu}$ denotes the viscous force from Shakura-Sunyaev viscosity \citep{ss73}. 
For boundary conditions, gas and dust densities, azimuthal velocities are fixed to initial values at inner and outer boundaries. Gas radial velocity is fixed to initial accretion velocity. Dust radial velocity at inner boundary is fixed to drift velocity whereas at outer boundary it is fixed to be zero such that there is no dust flowing into the disk at the outer disk boundary. This roughly fixes the total amount of dust inside our simulation domain.
As in our previous paper \citep{fu14}, the disk self-gravity and magnetic fields are neglected. The continuity equations of dust and gas are
\begin{equation}
\frac{\partial \Sigma_d}{\partial t} +\nabla \cdot (\Sigma_d \mathbf{v}_d) = \nabla \cdot \left(\Sigma_g D_d \nabla \left(\frac{\Sigma_d}{\Sigma_g}\right)\right)
\end{equation}
and
\begin{equation}
\frac{\partial \Sigma_g}{\partial t} + \nabla \cdot (\Sigma_g \mathbf{v}_g) =0,
\end{equation}
where $D_d$ is the dust diffusivity that is defined as $D_d=\nu_g/(1+\rm{St}^2)$ with gas viscosity $\nu_g$ \citep{takeuchi02}. 

We carried out simulations using our code 
\texttt{LA-COMPASS} (Los Alamos COMPutional Astrophysics Simulation Suite) in 2D cylindrical coordinates. 
In code units, the disk range is $r\in [0.4,6.68]$ and $\phi \in [0,2\pi]$. 
We apply observationally inferred parameters from the disk system in IRS48 \citep{marel13} for our model parameters. 
A 10 $M_{J}$ (Jupiter mass) 
planet is assumed to orbit a 2 $M_{\odot}$ (solar mass) star on a 
fixed circular orbit at radius $20$ AU which is taken as $r_p=1$ 
with Keplerian orbital frequency $\Omega_p$. The disk surface 
density at $r_p$ is  $\Sigma_g= 1$ which corresponds to $3\,\mathrm{g/cm^2}$. 
We also choose power-law profiles for initial disk surface density 
and temperature such that $\Sigma_g \propto r^{-1}$,  
$c_s^2 \propto r^{-0.5}$. The initial total mass in disk gas is $\sim 5 M_J$. 
The Shakura-Sunyaev viscosity $\alpha=7\times10^{-5}$ is set to be a 
constant throughout the disk. We choose $H|_{r_p}=0.06$ so the 
kinematic viscosity at the planet's orbit is $\nu|_{r_p}=2.5\times 10^{-7}$. 
The dust initial surface density distribution $\Sigma_d(r)$ follows $\Sigma_g(r)$. 
We have considered various initial surface density ratios $\eta_d =
\Sigma_d/\Sigma_g|_{t=0}$, ranging from $0.001$ to $0.01$. This implies that
the total mass of dust in the disk is $\sim 15 (\eta_d/0.01) M_{\oplus}$. 
The dust particle internal density is assumed to be $\rho_p=0.8\,\mathrm{g/cm^3}$. 
We have also considered different particle sizes, ranging from $10\, \mathrm{\mu m}$ to $4\,\mathrm{mm}$ with Stokes number spanning from $4\times 10^{-4}$ to $0.16$. Here the Stokes number is evaluated at $r=20$ AU using initial disk density profiles. For very small Stokes number (e.g. $4\times 10^{-4}$), we adopt short friction time approximation \citep{johansen05} to circumvent problems with small time-steps. However, this approximation is not valid for larger Stokes number (e.g. 0.01). For this reason, in each run we only model one particle species. Unless otherwise stated, runs in this study have grid resolution of $n_r \times n_{\phi}=6144\times 6144$ such that the planet Hill radius is resolved by 60 cells and a typical run lasts for 
several thousand planet orbits (one orbital period is $\sim 63$ yr).


\section{Results}

\begin{figure}
\epsscale{1.0}
\plotone{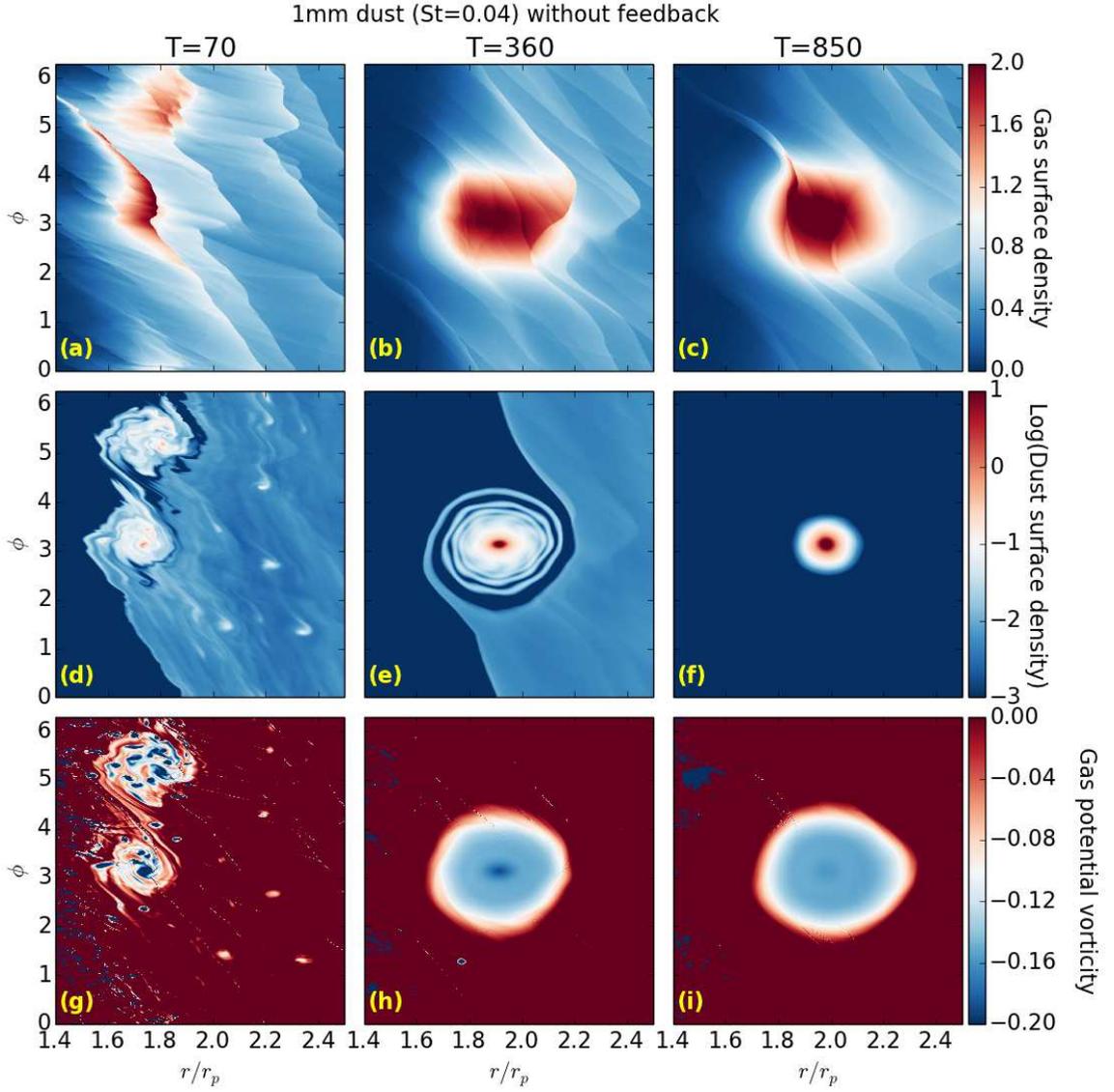}\\
\caption{Evolution of gas surface density (top row), log of dust surface density (middle row) and gas potential vorticity (bottom row) in the $\{ r, \phi\}$ plane. 
This run has 1 mm dust without any feedback effect and $\eta_d = 0.01$. (Color online)}
\label{fig:fig1}
\end{figure}

Figure \ref{fig:fig1} depicts the run without dust feedback. 
The gas disk and the 1 mm dust particles are evolved with $\eta_d=0.01$. 
Top, middle and bottom panels show the gas surface density, the logarithm of dust surface density and the gas potential vorticity perturbation (PV=$(\nabla\times\mathbf{\delta v}_g)_z/\Sigma_{g}$, i.e., initial potential vorticity subtracted), respectively. 
Each column represents a characteristic stage  
(in units of planet orbital period) of the evolution. 
At early stage, the planet carves out a clean gap whose outer edge 
(at $r\sim 1.6$) becomes RWI unstable and forms multiple Rossby 
vortices (panels [a], [d], [g]). These vortices quickly (within $\sim$ 
200 planet orbits) merge into a single vortex (panels [b], [e], [h]). 
This vortex acts as a very effective dust ``trap'' which collects dust 
particles that drift radially inward from the outer disk. 
Dust particles gradually accumulate at the vortex center and reach 
an even higher density than the gas component ($\rho_d \simeq 10$ in 
panel [e] v.s. $\rho_g \simeq 2.5$ in panel [b]). By the time T=850, 
the vortex has already taken in all the dust that originally 
resided in the outer disk (panel [f]). At this point the dust 
vortex is obviously much more confined and concentrated than the gas vortex.  

\begin{figure}
\epsscale{1.0}
\plotone{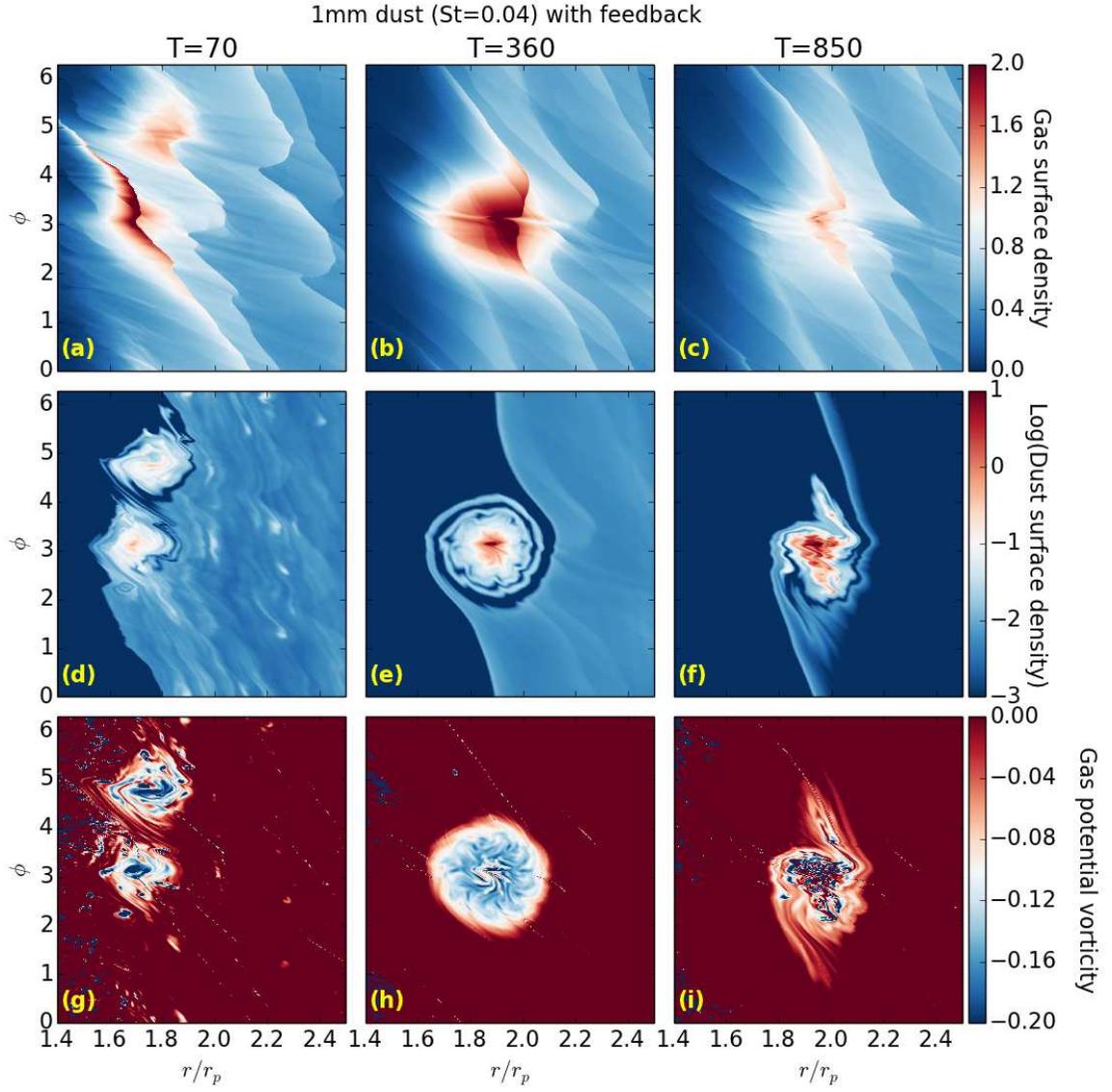}\\
\caption{Similar to Figure \ref{fig:fig1} except that dust feedback is included. (Color online)}
\label{fig:fig2}
\end{figure}

\begin{figure}
\centering
\includegraphics[width=1.0\textwidth]{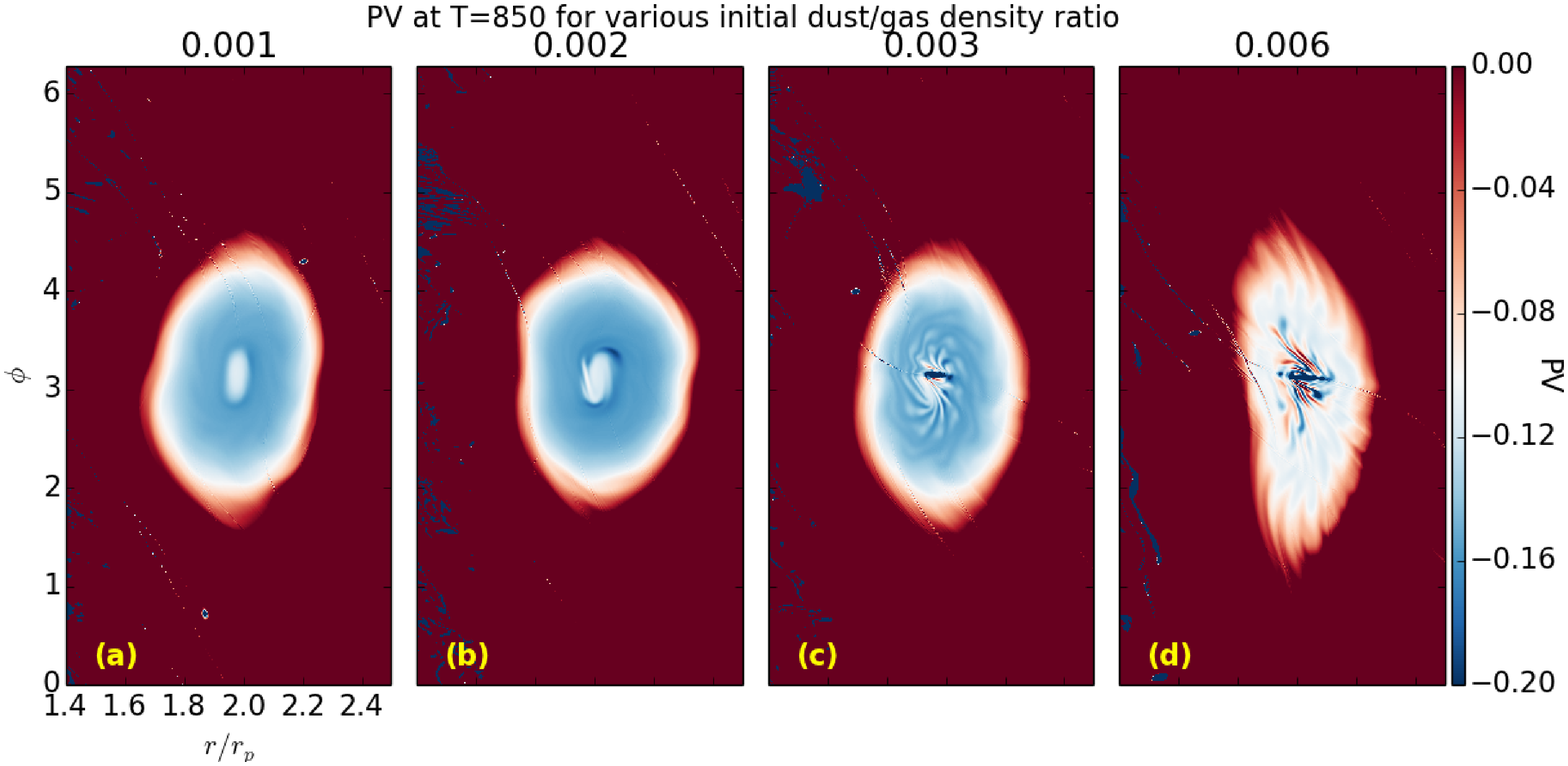}\\
\includegraphics[width=0.43\textwidth]{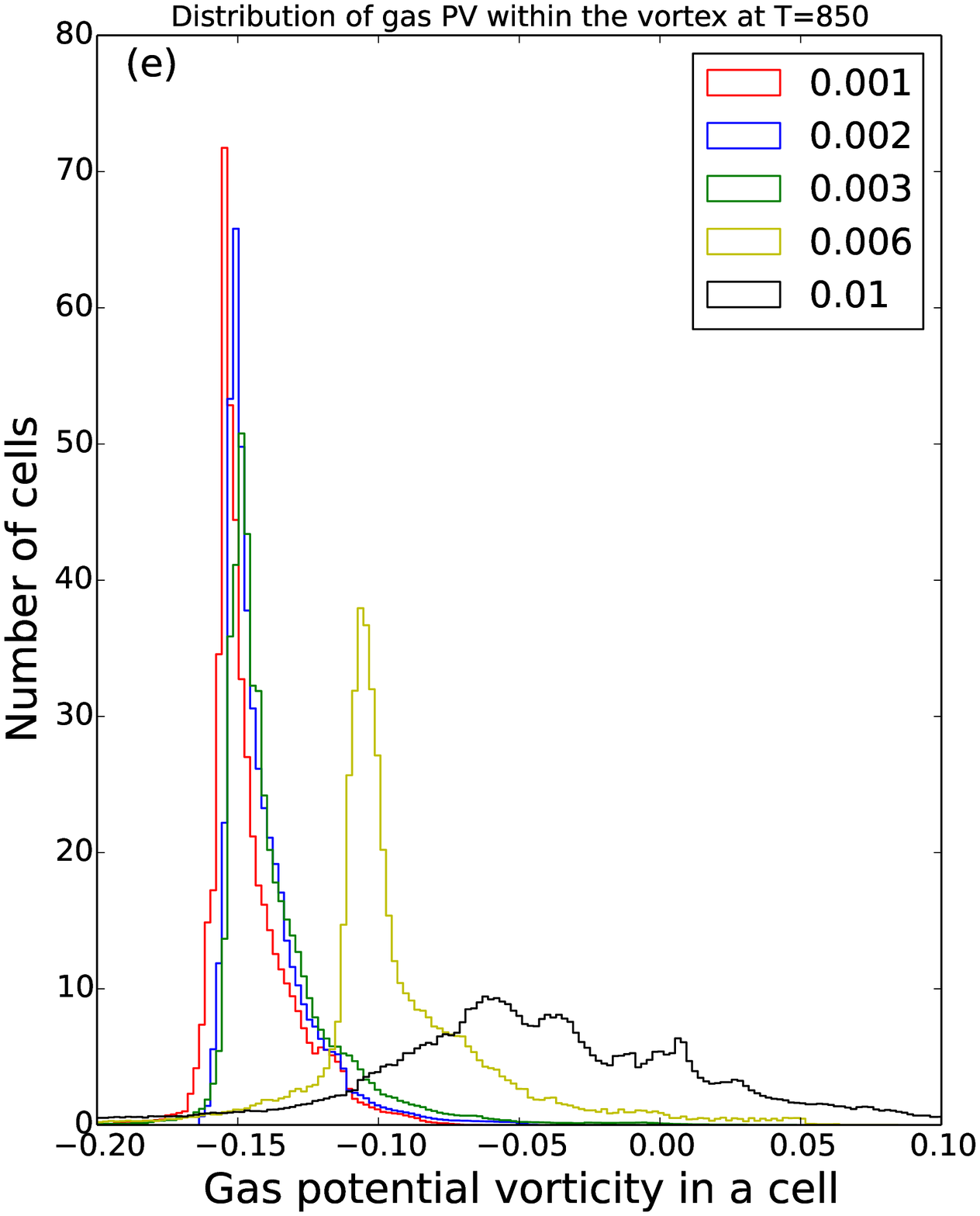}
\includegraphics[width=0.43\textwidth]{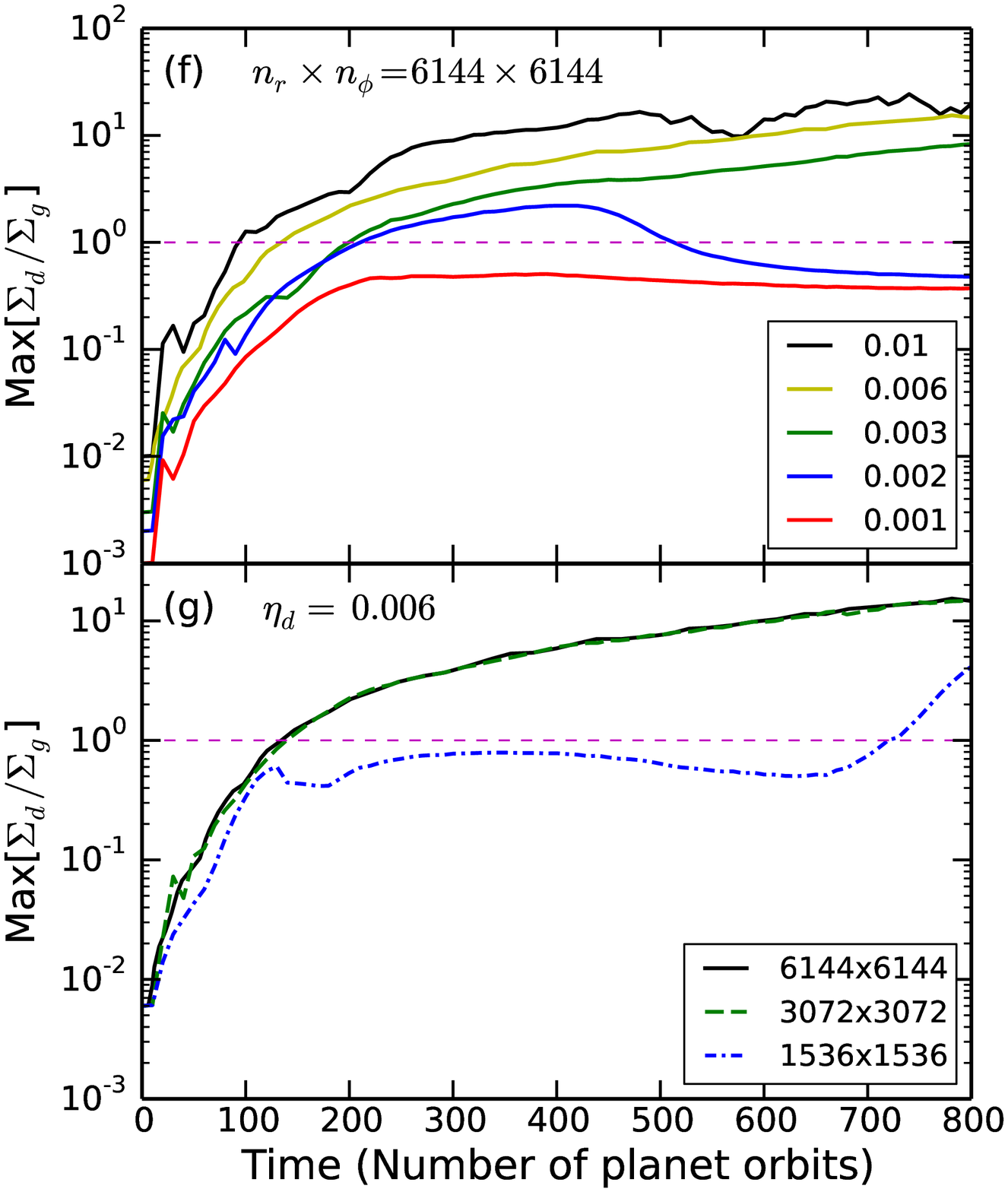}
\caption{Panels ({\it a-d}) show the gas potential vorticity at T=850 
for runs with various $\eta_d$. Panel ({\it e}) shows the histogram of 
gas potential vorticity within the vortex for based on panel ({\it a-d}). 
The evolution of maximal dust/gas density ratio within the vortex for various $\eta_d$ 
is shown in panel ({\it f}) and for various numerical resolutions 
(with $\eta_d = 0.006$) is shown in panel ({\it g}). All runs have 1 mm dust 
with feedback. (Color online)}
\label{fig:fig3}
\end{figure}

Figure \ref{fig:fig2} shows a run that has almost the same parameters 
as the run in Figure \ref{fig:fig1} except that dust feedback term is 
now included in the gas momentum equation. At T=850 the gas vortex (top row)
is significantly weakened, whereas the gas vortex in Figure \ref{fig:fig1}(c) 
has not yet started dissipating. Clearly dust feedback greatly accelerates the dissipation of the disk vortex in this case. The dust vortex also behaves 
differently. It starts developing a ridged surface at the vortex boundary and fluffy ``fingers'' around the core (panel [e]), instead of being smooth as in Figure \ref{fig:fig1}(e). It later becomes a clumpy collection of dust particles (panel [f]). Meanwhile, the gas potential vorticity distribution within the vortex also becomes very ``turbulent'' and irregular (panels [h], [i]), rather than having a smooth PV minimum as in Figure \ref{fig:fig1}(h)-(i). So far we have been using $\eta_d = 0.01$. 
In Figure \ref{fig:fig3} we considered four smaller initial density ratios. The comparison of their gas PV at T=850 (top panels) demonstrates that a higher $\eta_d$ 
leads to stronger feedback effects, thus faster vortex dissipation. Panel (e) of Figure \ref{fig:fig3} is the histogram of the gas PV of all the grid cells inside the vortex. We see that gas PV distribution both shifts to higher values and spreads out as 
$\eta_d$ increases. This quantitatively depicts the faster vortex dissipation with higher initial density ratio. The general rise of gas PV has to do with both surface density and vorticity change. However, the features we observe with high $\eta_d$ is mainly caused by vorticity pattern.

The features we see in the gas PV when dust feedback is included suggest that some type of dynamic instability is operating. This instability disrupts the smooth quasi-circular motion within the vortex and thus destroys the vortex. It operates only when the dust-to-gas feedback is included and when the dust/gas density ratio within
the vortex starts to reach unity or higher. In panel (f) of Figure \ref{fig:fig3}, we plot the maximal dust/gas density ratio inside the vortex as a function of time for various initial $\eta_d$. Obviously for higher $\eta_d$, larger dust concentration levels are achieved faster. The time point when a curve goes above unity is approximately when gas PV starts to exhibit features of dynamical instability. Note that for $\eta_d=0.002$, the blue curve bends over at T $\simeq$ 450 and falls below a value of unity at T $\simeq$ 500 due to dust dispersion. In panel (g) of Figure \ref{fig:fig3} we show the evolution of the maximal dust/gas density ratio for three different numerical resolutions with $\eta_d = 0.006$. 
Results from runs with relatively high resolution ($6144\times 6144$, $3072\times 3072$) differ only slightly, whereas the run with relatively low resolution ($1536\times 1536$; dot-dahsed line) significantly underestimates the concentration level of dust particles. The low resolution run could not even raise dust/gas density ratio up to a value of unity until about 600 orbits later. Thus it would have missed this instability completely if it is only run for less than 700 orbits.  
This result shows that high numerical resolution is needed in order to accurately capture the dust evolution inside disk vortices.

\begin{figure}
\epsscale{1.0}
\plotone{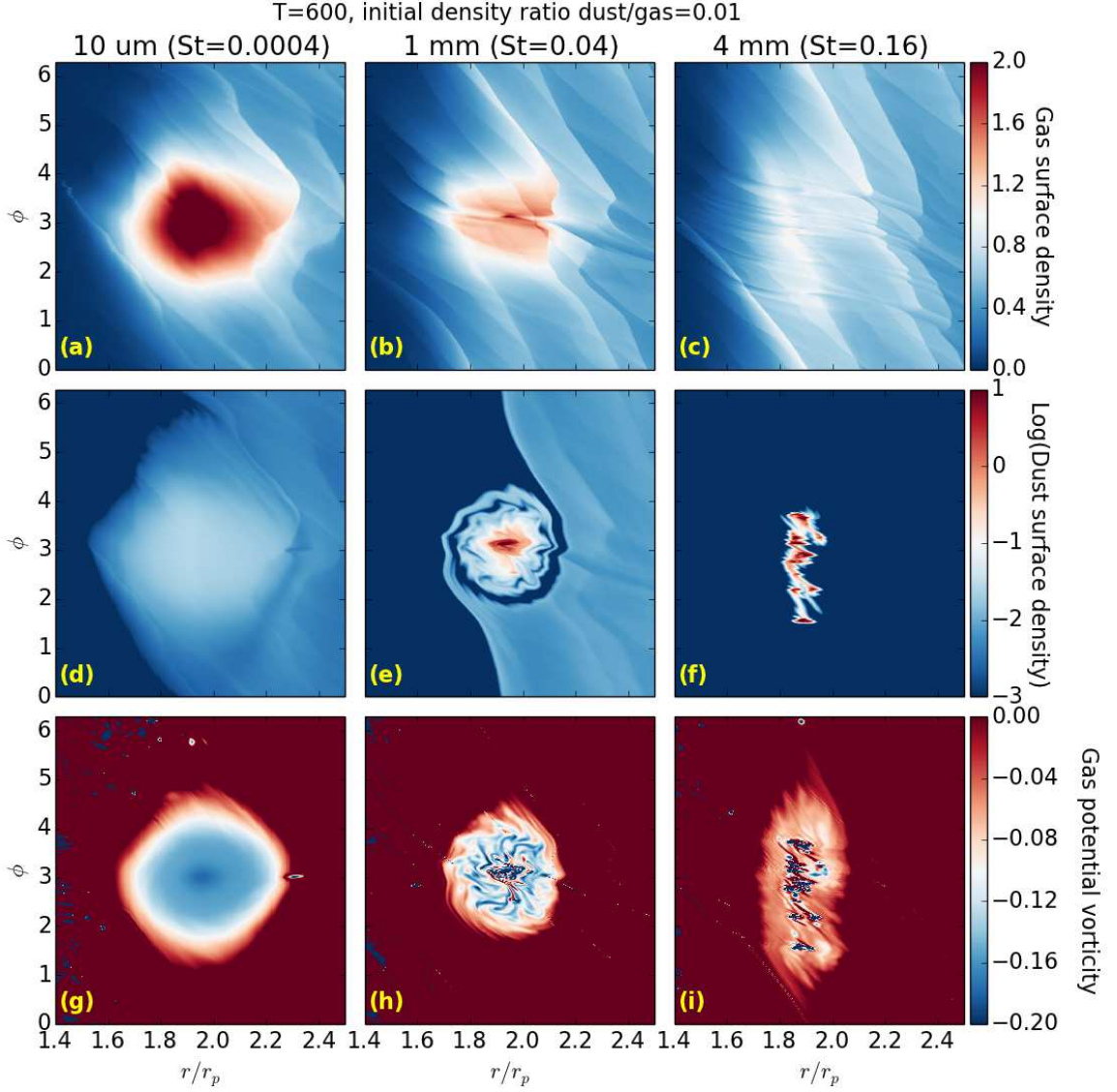}\\
\caption{Comparison of gas surface density (top row), log of dust surface density (middle row) and gas potential vorticity (bottom row) at T=600 for runs with dust sizes of 10 $\mathrm{\mu m}$, 1 mm, 4 mm, respectively. These runs all used $\eta_d = 0.01$ 
and included dust feedback. (Color online)}
\label{fig:fig4}
\end{figure}

Figure \ref{fig:fig4} shows a comparison of runs with different dust particle sizes, all with $\eta_d = 0.01$. We can see that the dust feedback effect becomes stronger with the 
dust particle sizes.

\begin{figure}
\centering
\includegraphics[width=0.8\textwidth]{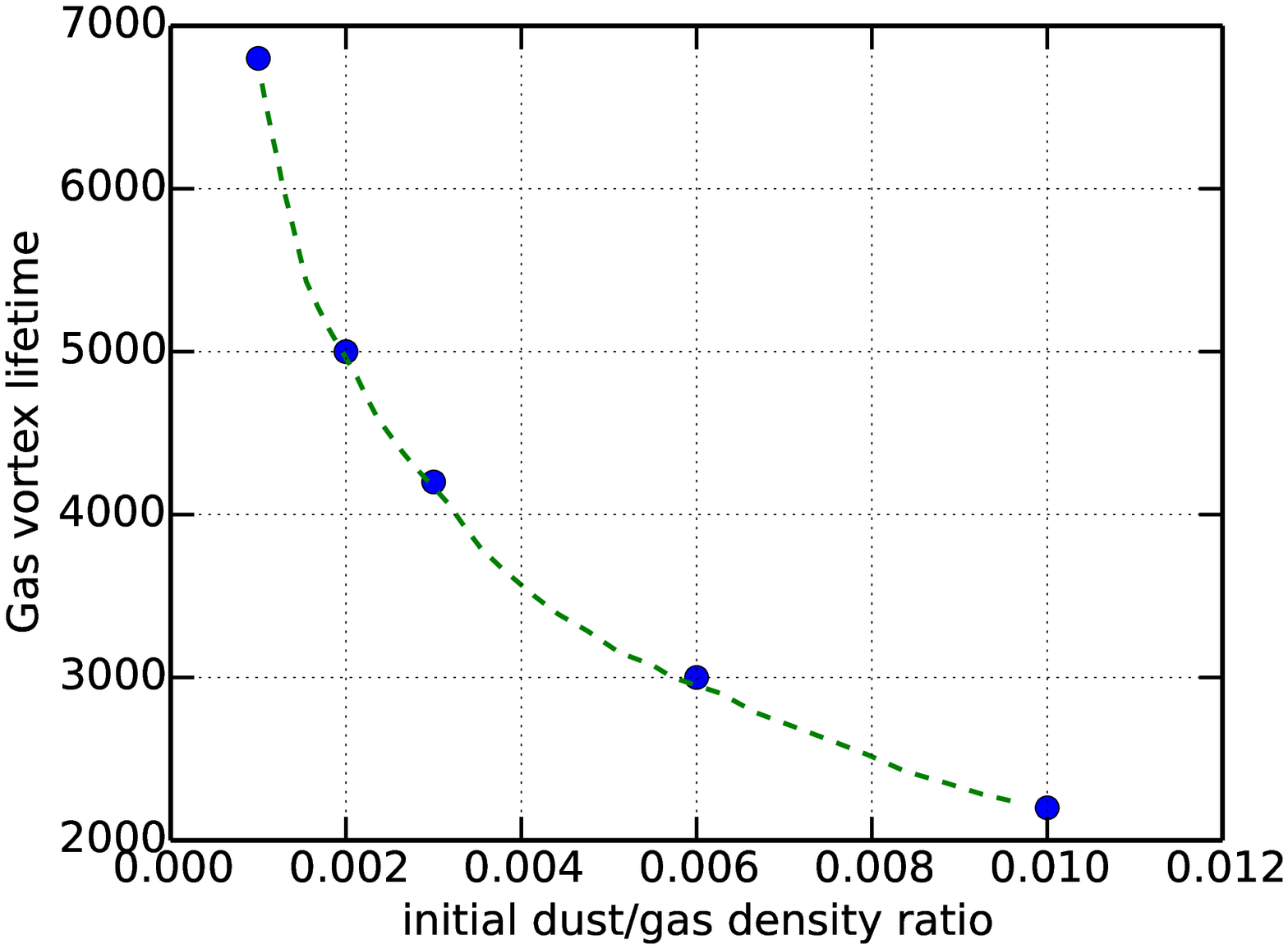}\\
\includegraphics[width=0.8\textwidth]{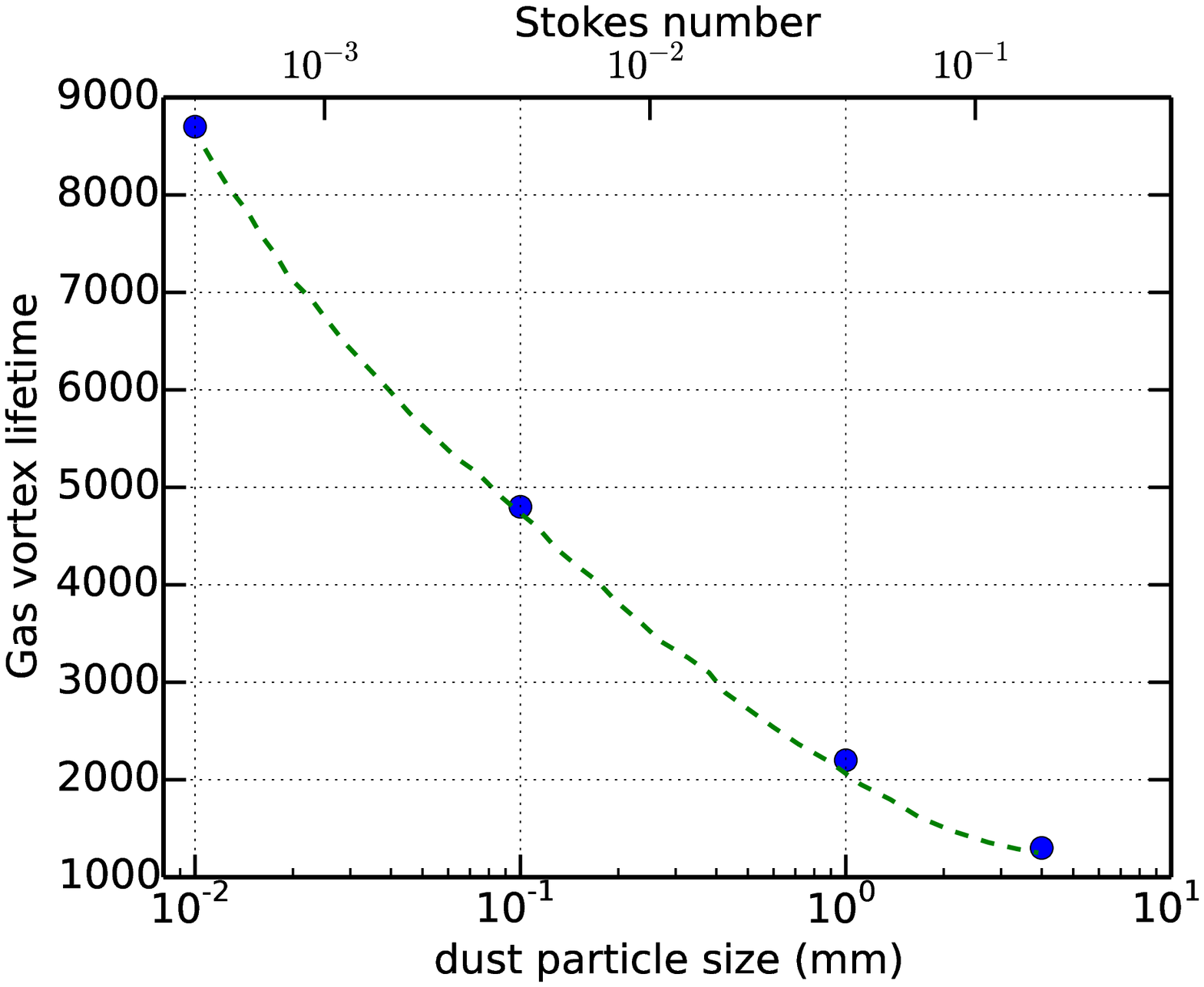}
\caption{Lifetime of gas vortices (in units of planet orbits) as a function of 
different $\eta_d$ with dust size of 1 mm (top) and different 
dust particle sizes with $\eta_d = 0.01$ (bottom). Dashed lines are rough interpolations. Dust feedback is included.}
\label{fig:fig5}
\end{figure}

The effect of dust feedback on the gas vortex lifetime is summarized in Figure \ref{fig:fig5} as a function of $\eta_d$ and dust particle sizes, respectively. 
The definition of vortex lifetime is the same as the one in our previous paper \citep{fu14}, i.e., a vortex is considered ``dead" when either the 
azimuthally averaged density variation or the azimuthally averaged  
potential vorticity variation within 10 H (scale height) wide band 
around the vortex drops below $10\%$. In order to run the simulation for 
a long enough time (close to $\sim$ $10^4$ orbits), we reduced the resolution in the runs ($n_r\times n_{\phi}=3072\times 3072$) for Figure \ref{fig:fig5}.

\section{Summary and discussion}

We have studied the back-reaction of dust trapping on planet-induced vortices in 2D protoplanetary disks. We found that disk vortices are very effective in collecting 
dust particles that are drifting inwards from the outer disk. For example,
with $\eta_d = 0.01$ and dust size of 1 mm, within 1000 orbits,
the disk vortex has accumulated up to 
almost $10 M_{\oplus}$ of dust (out of the total $15 M_{\oplus}$ available initially
in the disk). 
We also found that the dust-to-gas feedback can act to destroy the RWI vortices at the gap edge, i.e., shortening disk vortex lifetime. We have studied a range of the
initial dust to gas density ratio $\eta_d = 0.001 - 0.01$ and the dust particle
sizes with Stokes number $\sim 4\times10^{-4}$ to $0.16$ if we use 20 AU as the fiducial
radius based on the parameters of Oph IRS 48 \citep{marel13}. Within this parameter space, our simulations showed that higher $\eta_d$ and larger particle sizes lead to stronger feedback effects and thus shorter vortex lifetimes. Similar effects were also found in runs using different disk sound speeds. 

Within the vortex, dust feedback becomes quite important when 
the dust density is comparable to or higher than the gas density. 
The strength of dust feedback is therefore related to how fast dust converges 
in the vortex. If the dust mass within vortex grows slowly, then the vortex 
itself can be largely damped due to viscosity/shocks \citep[e.g.][]{fu14}. 
On the other hand, both higher initial dust/gas density ratio 
and larger dust grain size (faster drift velocity) give rise to more 
rapid dust accumulation within the vortex. These trends explain the 
dependence of disk vortex lifetime on these two parameters in Figure \ref{fig:fig5}.

When the dust density is high enough to affect flow dynamics, 
it excites some type of dynamical instability which modifies flow pattern within the vortex. This destroys the coherent, smooth potential vorticity minimum in the vortex (see Figure \ref{fig:fig2}(h), Figure \ref{fig:fig4}(g)). Since a local PV minimum is a necessary condition for RWI, when this condition is no longer satisfied, 
the vortex is not sustained and will dissipate quickly. \cite{johansen04} and \cite{meheut12} both carried out 3D simulations of dust-trapping vortices and reached similar conclusions as ours, i.e., the back-reaction of trapped dust could destroy a disk vortex. However, those two studies focused on only the very early evolution stage (up to several orbits) and their numerical resolution was relatively low. 
Our high resolution, long-term simulation, although in 2D, showed that the disk vortex gets destroyed not simply by dust dragging or splitting as those two papers reported, but also by a form of dynamical instability resulting from the complex dust-gas interaction. We are not certain about the physical nature of this instability. We know it is  neither the streaming instability \citep{youdin05, youdin07}, which only operates in 3D nor any other vortex instability that does not require dust-gas interaction (e.g. elliptical instability \citep{lesur09}).  It could be the heavy-core instability \citep{chang10} that was originally found in 2D analysis. However, we have not found a simple way to confirm this. Additional analyses are needed in order to pinpoint the nature of this instability which is beyond the scope of this \textit{Letter}.

In our simulations, each run includes just one particle species. In real disks, there is a distribution of particles with various sizes and densities. A more realistic modelling approach would be to include multiple particle species in one run. In that case, the dust feedback effect predominantly comes from the species which has the largest drift velocity and the highest surface density. Our study showed that the back-reaction from dust particles with Stokes number of only 0.16 (size of 4 mm in our disk model) could reduce the disk vortex lifetime by almost a factor of 10. Species with higher Stokes number (larger size) would make the effect more severe. It is already quite difficult to generate and sustain vortices in protoplanetary disks using conventional disk viscosity \citep{godon99, dvb07, ataiee13, fu14, zhu14}. To explain the observed disk asymmetries as planet-induced vortices, our finding thus provides constraints on the mass and distribution of dust particles. 

One interesting feature is that, with or without dust feedback, we found that the dust vortex/asymmetry almost always lives longer than gas vortex/asymmetry. Exactly how much longer depends on many system parameters. For the run in Figure \ref{fig:fig1}, 
after the disappearance of gas vortex, the dust concentration (panel (f)) will 
gradually spread out in the azimuthal direction and eventually become a 
uniform ring of dust at $r\sim 2$. The lifetime of the dust asymmetry is 
found to be about four times longer than the lifetime of the gas asymmetry 
in this case. When applied to the asymmetry in mm-wavelength dust emission of Oph IRS 48
and assuming that the mm (and even larger) particles are only a small portion of 
the total dust mass in order for dust feedback effect to be negligible,
the mm dust evolution can be well represented by the run in Figure \ref{fig:fig1}. What we see in ALMA images likely corresponds to a time of about a few thousand orbits. In this case, once gas vortex has disappeared, the asymmetry involving mm-sized dust is still strong, resulting in asymmetric dust emission at mm-wavelength that may correspond to the observed mm features. Smaller dust ($\mathrm{\mu m}$ sized) particles basically follow gas and we expect to see roughly uniform gas CO emission and dust emission at $\mathrm{\mu m}$ wavelength. However, our current model still has trouble explaining both the location (more than twice of the inferred planet orbit radius) and radial width (wider than what pressure could maintain) of the observed dust asymmetry in IRS 48.

Another feature that is worth noting from our simulations is that, given the high
dust concentration at the vortex center, we can estimate the effect of dust 
self-gravity by calculating the Toomre Q parameter of the dust. It comes to 
be around $1$ if we take the dust particle velocity dispersion is $1\%$ of the
Keplerian speed and the vortex 
dust surface density is $\sim 30$ g/cm$^2$ at $40$ AU. This suggests that dust
particles are likely to coagulate within the vortex and this should have very
interesting implications for planetesimal formation.

There are several limitations of our current investigation, though: i) Our study is 2D, whereas full dynamics can only be captured in 3D modelling \citep{meheut10}; ii) We adopted simple $\alpha$-viscosity, whereas real disk viscosity is likely produced by MRI turbulence; iii) We start the simulation with a limited supply of dust mass (limited disk size), whereas real disks could have a larger dust reservoir. Moreover, we ignored dust coagulation and fragmentation which are important parts of dust evolution especially when dust concentration level is high (e.g. at vortex center) \citep[see review of][]{testi14}. Nevertheless, \cite{zhu14} recently found that vortex modelling in 2D viscous disks seemed to reproduce results obtained from unstratified 3D MRI-disks. In addition,
previous 3D studies \citep{johansen04, meheut12} have also pointed out the adverse
influence of dust on vortices. Therefore, we expect that our general conclusion still holds in more realistic situations, i.e., dust feedback makes it more difficult to sustain vortices in protoplanetary disks. 


\section*{Acknowledgements}
Simulations in this work were performed using the Institutional Computing Facilities 
at LANL. We thank Geoffroy Lesur and Andrea Isella for helpful discussions. WF, HL and SL gratefully acknowledge the support by the LDRD, IGPP 
programs, UC laboratory fees research program and DOE/Office of Fusion Energy Science through CMSO at LANL.  
WF and SL also acknowledge support from NASA grant NNX11AK61G.


\end{document}